\documentstyle[eclepsf,epsf,wrapft]{ptptex}

\preprintnumber{HUPS-96-2}

\title{
On the asymptotic equations 
in the harmonic oscillator representation
}

\author{
Gennady F. {\sc Filippov}$^{1}$, 
Alexei D. {\sc Bazavov}$^{2}$, \\
Kiyoshi {\sc Kat\=o}$^{3}$, 
and Sergei V. {\sc Korennov}$^{1,3}$
}

\inst{
$^{1}$ Bogoliubov Institute for Theoretical Physics, 
National Acad. of Sci., \\ Kyiv, 252143 Ukraine \\
{\rm E-mail: gfilippov@gluk.apc.org}
$^{2}$ National University of Ukraine, Kyiv, Ukraine \\ 
$^{3}$ Graduate School of Science, Hokkaido University, 
Sapporo 060, Japan\\
{\rm E-mail: kato@nucl.phys.hokudai.ac.jp, 
korennov@nucl.phys.hokudai.ac.jp}
}

\recdate{
\today
}

\abst{
In the harmonic oscillator representation, the Schr\"odinger equation
has a form of a set of infinite number of algebraical equations
which are labeled by the radial quantum number $n$. It is shown that
at $n \gg 1$ these  equations are significantly simplified
and become an algebraical analogue of the original differential 
Schr\"odinger equation. This result is generalized to the case
of multi-channel systems being studied in the framework of the 
resonating group method.
}

\begin{document}

\maketitle

\section{Introduction}
The Schr\"odinger equation for a particle moving in a spherical-symmetry
field 
$U(r)$,
\begin{equation}
\label{eq-1}
\{T+U(r)\}\Psi({\bf{r}})=E\Psi({\bf{r}}),
\end{equation}
can be transformed to 
the form 
of an infinite set of algebraical equations 
for the Fourier coefficients
of the harmonic oscillator expansion of the wave function 
$\Psi({\bf{r}})$;
\begin{equation}
\sum_{\tilde{n}=n-1}^{n+1}<nl|T|\tilde{n}l>C_{\tilde{n}l}+
\sum_{\tilde{n}=0}^{\infty}<nl|U(r)|\tilde{n}l>C_{\tilde{n}l} -EC_{nl}=0,
~~n,l=0,1,2,...
\label{eq-2}
\end{equation}
\begin{equation}
\Psi({\bf{r}})=\sum_{lm}R_l(r)Y_{lm}(\Omega),~~R_l(r)=\sum_{n=0}^{\infty}
C_{nl}\Phi_{nl}(r),
\end{equation}
\begin{equation}
\Phi_{nl}(r)=(-1)^n\sqrt{{2\Gamma(n+1)\over{\Gamma(n+l+3/2)}}}
r^lL_n^{l+1/2}(r^2)\exp\{-r^2/2\},
\end{equation}
where, as usually, $l$ is the orbital momentum, $n$ is the number of the radial 
quanta, and $L_n^{l+1/2}(r^2)$ are the Laguerre polynomials. The oscillator length
$r_0$ is set to unity.

The set of this form are obtained after the realization of the multi-channel 
resonating group method and the hyperspherical function method in the 
harmonic oscillator basis.

Note that even in the case of  the local potential $U(r)$ the set of its
matrix elements
\begin{eqnarray*}
\{<nl|U(r)|\tilde{n}l>, 0\leq n,l < \infty\}
\end{eqnarray*}
is actually a discrete analogue of a non-local potential, so that the investigation
of the infinite set (\ref{eq-2}) is complicated. In Refs.[1-3], it was shown how the 
set (\ref{eq-2}) can be completed by using the quasi-classical approximation for the 
Fourier coefficients, where one can restrict himself with a finite number
of equations. However, if one deals with slowly decreasing potentials with power-like
behavior at the infinity, the problem of the asymptotic
Fourier coefficients remains, and for solving this problem, 
it is necessary to carry out 
an additional investigation.

In this paper, the asymptotic equation in the harmonic oscillator
representation is discussed and the asymptotic form of the potential
is derived. With the use of the results,
we can obtain a knowledge of the asymptotic behavior of a potential
from its matrix elements. Although usually the potential itself is
known, it is very important to apply the present method to many-body systems
described with the use of a two-body potential. We can know the 
asymptotic behavior of the hyperradial potential even if we do not
know its explicit form$^4$. Information about the asymptotic
wave function including the influence of long-range properties 
of a potential becomes very important in the complex scaling 
calculations$^5$.   

\section{Derivation of asymptotic equations}

Let us recall how the problem of the asymptotic wave function 
$\Psi({\bf{r}})$
has been solved. In Eq.(\ref{eq-1}) with a limitation to large 
$r$, where the behavior of the potential $U(r)$ becomes simpler because, in its
expansion in the powers of $r$, the only main term remains, and then the precise
solution of a rather simple asymptotical equation is found.

In the case of the set ({\ref{eq-2}), the number of radial quanta $n$ 
is well analogous to
 the radius $r$. Therefore, we shall try to reduce those equations in
(\ref{eq-2}) which correspond to $n \gg 1$. It is known$^6$  that at 
$n \gg 1$, the 
discrete variable $\sqrt{4n+2l+3}$ can be replaced by a continuous one,
$r$. Then, the first summation in Eq.(2),
$$
<nl|T|n-1,l>C_{n-1,l} + <nl|T|nl>C_{nl} + <nl|T|n+1,l>C_{n+1,l}
$$
appears to be an algebraical analogue of the differential 
kinetic energy operator 
$T$ multiplicated by $1/\sqrt{r}$, i.e.
$$
{1\over\sqrt{ r}} \left\{ - {1\over r^2}{d\over dr}{r^2}{d\over dr}
+ {l(l+1)\over r^2} \right\} = {1\over\sqrt{r}} T,
$$
acting onto the wave function
$$
R_{l} = R_l(\sqrt{4n+2l+3}) = {C_{nl} \over \sqrt2 (4n+2l+3)^{1/4}} .
$$
Therefore, we are mainly interested in the summation
\begin{equation}
\label{eq-5}
\sum_{\tilde{n}=0}^{\infty} <nl|U(r)|\tilde{n}l>C_{\tilde{n}l} .
\end{equation}
In this summations, the terms have the following explicit form;
\begin{equation}
\int_0^\infty \Phi_{nl}(r)U(r)\Phi_{\tilde{n}l}(r)r^2 dr
\int_0^\infty R_l(\tilde{r})\Phi_{\tilde{n}l}(\tilde{r}){\tilde{r}}^2 d\tilde{r}.
\end{equation}
The basis of the radial functions $\Phi_{nl}(r)$ satisfies the condition
\begin{equation}
\sum_{\tilde{n}=0}^{\infty} \Phi_{{\tilde{n}}l}(r)r
\Phi_{{\tilde{n}}l}(\tilde{r})\tilde{r} = \delta(r-\tilde{r}).
\end{equation}
Having changed the order of operations in (5-6) and performed 
summation, we have
\begin{eqnarray*}
\sum_{\tilde{n}=0}^{\infty} <nl|U(r)|\tilde{n}l>C_{\tilde{n}l} =
\int_0^\infty \int_0^\infty \Phi_{nl}(r)U(r)R_l(\tilde{r})
\delta(r-\tilde{r}) rdr \tilde{r} d\tilde{r}=
\end{eqnarray*}
\begin{equation}
\label{eq-8}
=\int_0^\infty \Phi_{nl}(r)U(r)R_l(r)r^2 dr.
\end{equation}

The basis functions $\Phi_{nl}(r)$ have a remarkable asymptotic behavior.
That is, if 
$n\gg1$, then
\begin{equation}
\label{eq-9}
\Phi_{nl}(r)r^{3/2} \simeq \sqrt{2} \delta(r-r_n),
\end{equation}
where $r_n=\sqrt{4n+2l+3}$ is a turning point in the field $r^2/2$ 
of the harmonic oscillator for the particle moving with the energy 
$2n+l+3/2$. As a result, the last integration in (\ref{eq-8}) can be performed
at $n\gg1$, and the formula (\ref{eq-5}) is significantly simplified,
\begin{equation}
\sum_{\tilde{n}=0}^{\infty} <nl|U(r)|\tilde{n}l>C_{\tilde{n}l} \simeq
U(\sqrt{4n+2l+3})\sqrt{2r_n}R_l(r_n) \simeq U(\sqrt{4n+2l+3}) C_{nl}.
\end{equation}
Thus, in the limit of large $n$, the following asymptotic form of the set 
(\ref{eq-2}) is valid;	
\begin{equation}
\label{eq-11}
\sum_{\tilde{n}=n-1}^{n+1} <nl|T|\tilde{n}l>C_{\tilde{n}l}+
U(\sqrt{4n+2l+3})C_{nl} -EC_{nl}=0.
\end{equation}
Bearing in mind the remark concerning the first sum in (11), we come
to a conclusion that Eq.(11) does not differ from the asymptotic form of the 
Schr\"odinger equation (1) at large $r$, i.e., from (11) we can turn back 
to (1) with a local potential $U(r)$ but only if $r \gg 1$.

The set (\ref{eq-11}) is a three-term recurrent relation in 
$C_{n+1,l}$, $C_{n,l}$, and $C_{n-1,l}$. If the limit expression
for $R_l(r)$ at $r\gg1$ is known, then $C_{n,l}$ may be calculated due to 
the equality
\begin{equation}
C_{n,l}=\sqrt{2r_n} R_l(r_n).
\end{equation}
After that, the recurrent relation (\ref{eq-11}) allows us to go backward
to that values of $n$ at which it remains valid.

Having chosen this way of completing the set (\ref{eq-2}), in the result we 
will have only small number of equations to be solved exactly. 

\section{On the convergence to the asymptotical limit}
\begin{wrapfigure}{l}{7.25cm}
 \epsfxsize=8cm
 \centerline{\epsffile{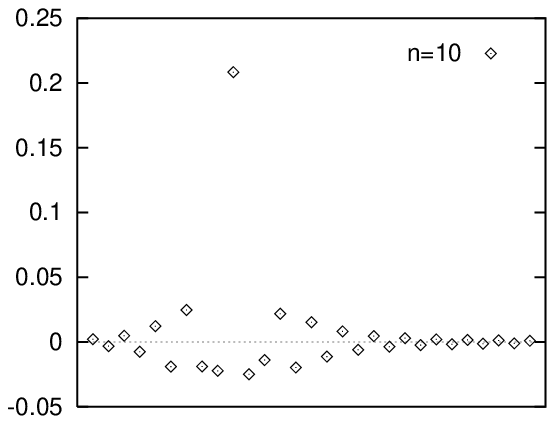}}
 \vspace{-6.85mm}
 \epsfxsize=8.0cm
 \centerline{\hspace{0mm}\epsffile{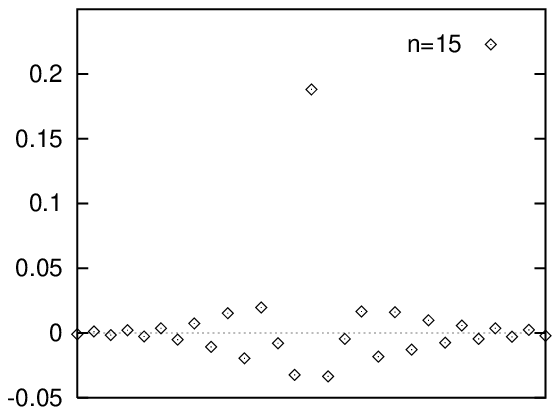}}
\vspace{-6.65mm}
 \epsfxsize=8cm
\centerline{\epsffile{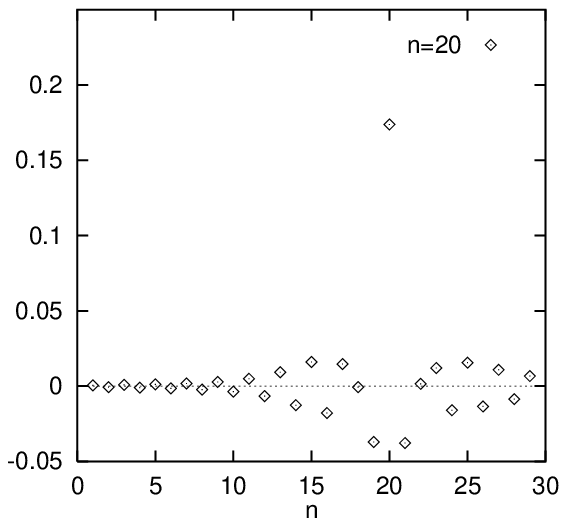}}
\caption{
Matrix elements $<nl|U(r)|\tilde{n}l>$
of the potential (14) with $\gamma=0.25$. 
}
\end{wrapfigure} 

Now we turn to the question; for which values of $n$ the 
asymptotic formula (\ref{eq-9}) is valid? Actually, this formula means 
that,
at large $n$, the set of matrix elements 
$<nl|U(r)|\tilde{n}l>$ is a discrete analogue of the $\delta$-function 
with the normalization factor 
$U(\sqrt{4n+2l+3})$. The matrix elements 
themselves do not equal to zero only in the vicinity of the diagonal
$\tilde{n}=n$, and they oscillate when the value of $\tilde{n}-n$ 
is changed (their signs depend on the parity of  
$\tilde{n}$), with the oscillation amplitude decreasing to zero. If the values
of  
$U(r_n)$ are not known {\it a priori}, they can be found by performing the 
summation of $<nl|U(r)|\tilde{n}l>$:
\begin{eqnarray}
\sum_{\tilde{n}=0}^{\infty} <nl|U(r)|\tilde{n}l>  \nonumber
\\
= U(\sqrt{4n+2l+3}).
\end{eqnarray}
In the summation over $\tilde{n}$, the upper limit must be not less 
than $2n$. 
Then it is enough to take $n>20$ in order to make the sum converge.

Figure 1 illustrates the behavior of the matrix elements 
$<nl|U(r)|\tilde{n}l>$ at different $n$ for the case
\begin{equation}
\label{pot-1}
U(r) = {1 - \exp(-\gamma r^2) \over r}, 
\end{equation}
while Fig.2 (left) shows the values of the sums (13) for this potential.
Convergence of the sums are observed at $n=20$ and even a little less.

Another example is a potential 
\begin{eqnarray}
\label{pot-2}
U(r)={1\over r^3} \left( 1 - \gamma r^2 
-  \exp(-\gamma r^2) \right),
\end{eqnarray}
which decreases as $1/r^3$ and by this reason it is characteristic
of the hyperspherical function method\cite{jmp-1}. The behavior of the sums (13)
of the matrix elements of this potential is shown in Fig.2 (right). 
Again,
the sums are rapidly converge to the limit 
value.

All said above remains valid for the potentials which are not singular in 
the origin.

\begin{wrapfigure}{c}{14cm}
 \epsfxsize=7.25cm
 \centerline{
 \epsffile{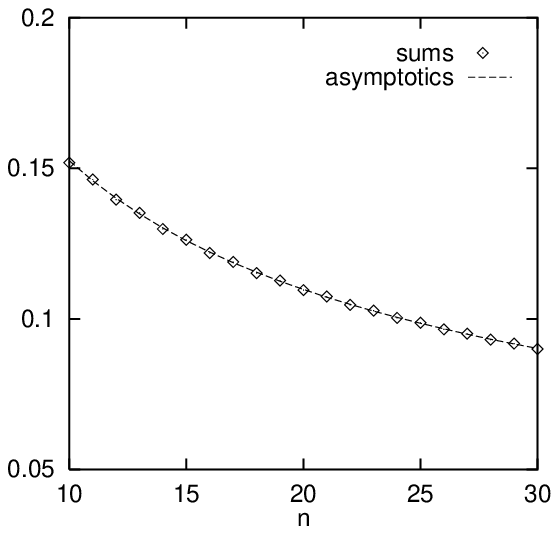}
 \epsfxsize=7.25cm
\epsffile{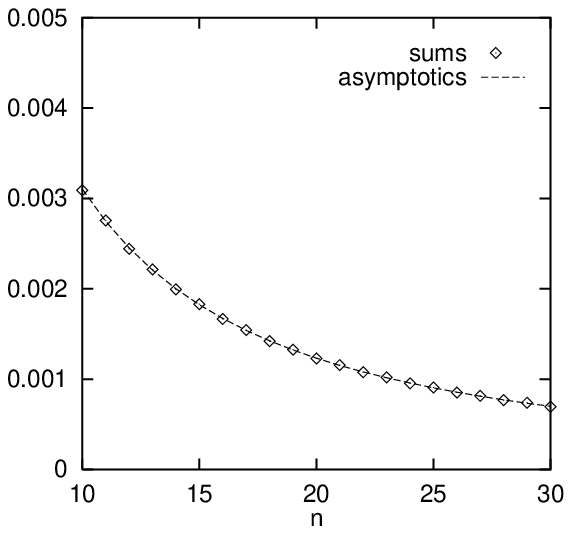}}
\caption{
Asymptotics and the sums (13) 
of the potentials (14) (left) and (15) (right) with $\gamma=0.25$.
}
\end{wrapfigure} 
%
%

\section{Generalization}

For a many-channel system being considered within the frame of 
the algebraic version of the resonating group method, when the 
reactions with light nuclei are investigated, the harmonic
oscillator basis states $|\alpha, nl>$ have an additional index 
$\alpha$ for distinguishing between the channels. The Fourier 
coefficients accept the same index. If the basis states of different
channels are coupled only by the operator of the potential energy, 
then instead of Eq.(11) we get
\begin{equation}
\label{eq-14}
\sum_{\tilde{n}=n-1}^{n+1} <\alpha, nl|T|\alpha, \tilde{n}l>
C^\alpha_{\tilde{n}l} +
\sum_{\beta=1}^p U_{\alpha\beta} (\sqrt{4n+2l+3})C^\beta_{nl}
- EC^\alpha_{nl} = 0 ,
\end{equation}
$$
\alpha = 1,2, \dots , p ,
$$
where $p$ is a number of channels. In other words, in this case 
there appear asymptotic matrix elements of the potential energy
of the interaction between channels, and they contribute to 
the asymptotics of the Fourier coefficients $C^\alpha_{nl}$.

A practically inportant point (especially, in the low-energy region)
is consideration of the Coulumb repulsion of the 
charged particles (e.g. protons) which
leads to the polarization of clusters. Then, the matrix elements 
$U_{\alpha\beta}$  decrease with $n$ increases,
\begin{equation}
\label{eq-15}
U_{\alpha\beta} \simeq {A_{\alpha\beta} \over \sqrt{4n+2l+3}},
\end{equation}
and it is necessary to trace over the influence of the Coulumb 
interaction between channels starting at very large $n$.

The algebraic version of the hyperspherical function method
employs the basis of the states $|\alpha, nK>$ where $K$ is the 
hypermomentum and $n$ includes the remaining quantum numbers.
The asymptotical equations for the hyperspherical function methods 
do not differ in principle from Eqs.(\ref{eq-14}). However, $K$ appears
instead of $l$, and $\sqrt{4n+2l+3}$ must be replaced by 
$\sqrt{4n+2K+3(N-1)}$ where $3(N-1)$ is the dimension of the
space in which the hyperharmonics are defined.

\section{Conclusion}

In the harmonic oscillator representation, after the matrix
elements of the Hamiltonian $<nl|T+U|\tilde{n}l>$ ($n$ is the
radial quantum number, $l$ is the orbital momentum) are found,
the Schr\"odinger equation has a form of a set of algebraical
equations in the unknown Fourier coefficients $C_{nl}$. In the
limit of great $n$ ($n \ge 20$), these equations are reduced
and become a discrete analogue of the Schr\"odinger equation
at large values of the radius $r$. Therefore, there is a 
one-to-one correspondence between the asymptotic 
radial function $R_l(r)$ at large $r$ and the asymptotic 
Fourier coefficients $C_{nl}$ at large $n$. This result
can be generalized to the case of many-channel systems.

\section*{Acknowledgements}
The authors would like to acknowledge the following organizations 
for supporting this research;

The Ministry of Education, Science and Culture of Japan

Japan Society for Promotion of Science

International Science Foundation

INTAS

Ukrainian Committee of Science and Technology.

\end{document}